# Vibrational coupling to quasi-bound states in the continuum under tailored coupling conditions

Keisuke Watanabe[1]*, Hemam Rachna Devi[2], Masanobu Iwanaga[3], and Tadaaki Nagao[2,4]

1.* International Center for Young Scientists (ICYS), National Institute for Materials Science (NIMS), 1-1 Namiki, Tsukuba, Ibaraki 305-0044, Japan, E-mail: watanabe.keisuke@nims.go.jp. https://orcid.org/0000-0002-4285-2135
2. International Center for Materials Nanoarchitectonics (MANA), National Institute for Materials Science (NIMS), 1-1 Namiki, Tsukuba, Ibaraki 305-0044, Japan.
3. Research Center for Electronic and Optical Materials, National Institute for Materials Science (NIMS), 1-1 Namiki, Tsukuba, Ibaraki 305-0044, Japan.
4. Department of Condensed Matter Physics, Graduate School of Science, Hokkaido University, Kita 10, Nishi 8, Kita-ku, Sapporo 060-0810, Japan.

**Abstract**
Photonic resonance modes can be spectrally coupled to the vibrational modes of molecules in the mid-infrared regime through interactions between localized electric fields and nearby molecules. According to recent studies, radiative loss engineering of coupled systems is a promising approach for tailoring coupling conditions and enhancing the molecular signals. However, this strategy has only been realized using the localized surface plasmon resonances of metal nanostructures, which suffer from increased ohmic loss in the mid-infrared region and face serious limitations in achieving high quality ($Q$) factors. In this study, we adopt silicon-based metasurfaces formed on silicon-on-insulator wafers to achieve high $Q$ factors and tune the coupling conditions between the quasi-bound states in the continuum (qBICs) and molecular vibrations. The coupling between the resonance mode and polymethyl methacrylate molecules is tailored from weak to strong coupling regimes by simply changing the structural asymmetry parameter and utilizing the intrinsically high $Q$ factors of the qBIC modes. In addition, we identify the optimal asymmetry parameter that maximizes the enhanced molecular signal, opening a route toward realizing highly sensitive surface-enhanced infrared spectroscopy using complementary metal-oxide semiconductor compatible all-dielectric materials.

**Introduction**
Resonantly coupled systems between photonic resonators and molecules are characterized by spectral variations resulting from light–matter interactions [1]. Photonic nanostructures are advantageous owing to their strong near-field confinement at the nanoscale, which enables strong interactions with nearby molecules [2]. If the coupling rates between the resonators and molecules are sufficiently higher than the constituent damping rates, the coupled system reaches a strong coupling regime and forms a newly hybridized or polariton state [3], exhibiting mode splitting in the spectra. Compared to coupling with electronic transitions [4–6], which has attracted significant attention over the past decade, vibrational transitions have much smaller transition dipole moments [7]. Nevertheless, by utilizing an enhanced energy exchange rate between the collective dipole resonance of high-density molecular vibrations and the single resonance mode [8], strong coupling between plasmonic nanostructures and molecular vibrations [9–12], including polymethyl methacrylate (PMMA) molecules [13, 14], has been reported quite recently. The key challenge in achieving a strong coupling regime is to reduce the damping rate of the nanostructure compared to the coupling rate $g$ with the molecules. Thus, it is important to precisely control the losses in the structure. However, an elaborate structural design is required, and the variable range of losses is limited by the large ohmic loss and collisional damping of metals. In this respect, dielectric nanostructures have advantages in terms of low material losses and wide tunability of radiative quality ($Q$) factors [15]. In particular, dielectric metasurfaces at quasi-bound states in the continuum (qBICs) can control their radiative losses simply by changing their asymmetry parameters [16, 17], thus offering great potential for the extensive tuning of the coupled system from weak to strong coupling regimes.

Another advantage of loss engineering is that it favors molecular sensing based on surface-enhanced infrared absorption (SEIRA) spectroscopy. Identification of a small number of molecules in the mid-infrared range has conventionally been realized by utilizing the interactions between molecular vibrations and strongly localized electric fields in plasmonic platforms [18, 19]. Recently, loss engineering for maximizing molecular absorption at the vibrational modes has been reported as another strategy [20–24]. The ratio of radiative to nonradiative intrinsic loss rates originating from metallic nanostructures can be controlled by changing the structural parameters, resulting in a large vibrational signal under optimum conditions [25, 26]. This method provides significant enhancement without requiring precise nanofabrication, such as a few nanometer gaps [23]. Despite their highly desirable features, the feasibility of enhancing molecular signal strengths with all-dielectric materials has not yet been explored and demonstrated.

In this study, we theoretically and experimentally demonstrate the enhancement of molecular signal strength by engineering the coupling conditions between the qBIC modes of silicon metasurfaces and the vibrations of PMMA molecules. The structural asymmetry parameters of the asymmetric pair-rod arrays are selected such that the Q factors of the qBIC modes varies independently while maintaining the resonance wavelengths. The coupling conditions of the qBIC-PMMA system are readily tuned from weak to strong coupling regimes by simply changing the asymmetry parameters, whereas the upper limits of the Q factors are determined by the leakage losses into the high-index substrate of the silicon-on-insulator (SOI) wafer. We also show the existence of an optimum asymmetry parameter that maximizes the molecular signal, as explained by temporal coupled mode theory. These conditions are significantly affected by the downward leakage losses when the oxide layer thickness of the SOI wafer is changed. To the best of our knowledge, this is the first report to show that the loss engineering of all-dielectric metasurfaces can increase the molecular vibrational signals, providing a means for highly sensitive surface-enhanced infrared spectroscopy without using metals.

**Results and Discussion**
**1. Structure and characteristics**
Figure 1a shows the proposed silicon metasurface with asymmetric pair-rod arrays on an SOI wafer with 400-nm thick silicon and a 2000-nm thick buried oxide (BOX) layer. The period of array $P$ is 3900 nm, the primitive silicon rod length $L$ is 2625 nm, the rods height $w$ is 985 nm, and the distance between the centers of the pair-rod is 1825 nm. The qBIC condition arises from symmetry breaking of the unit structure of the metasurface that satisfies the BIC condition with ideally infinite $Q$ factors, even inside the light cone [27]. In our study, asymmetry is added by varying the lengths of the upper and lower rods in the unit cell, whose asymmetry parameter is expressed as:

$$\alpha = \frac{2\Delta L}{L}. \tag{1}$$

Figure 1b shows a representative scanning electron microscopy (SEM) image of a metasurface fabricated using electron-beam lithography and deep silicon etching (see Methods). As an example, we simulate the field strength $|E|^2$ at the qBIC resonance wavelength ($\alpha$ = 0.2) for the $xy$-plane at half the height of the silicon rods ($\alpha$ = 0.2) using the finite-difference time-domain (FDTD) method. Figure 1c shows the strongly localized electric field intensity at the sidewalls of the silicon rods, which induces a strong interaction with the coupled molecules. The white arrows indicate the existence of anti-parallel displacement currents in the upper and lower rods, which give rise to a net electric component in the $x$-direction. Consequently, the qBIC resonance mode (nonzero $\alpha$) can be excited in free-space using a normally incident $x$-polarized plane wave. We also calculate the spatially averaged $|E|^2$ spectrum ($\alpha$ = 0.2) for the $xy$-plane at the center of the silicon nanostructures (lower panel on Fig. 1c). At the peak wavelength, the near-field intensity is resonantly enhanced by the qBIC resonance.

The distinctive feature of the asymmetric pair-rod structure is that the linewidths can be independently tuned, while the resonance peak wavelengths remain almost constant when varying $\alpha$, which allows for easy control of the coupling conditions with molecular vibrations. Figure 1d shows the simulated transmission spectra of bare silicon metasurfaces with different $\alpha$. Both the resonance peak amplitudes and linewidths increase with parameter α while maintaining the peak positions. In the experiment, the transmission spectra of a 3 mm × 3 mm sample were acquired at normal incidence using Fourier-transform infrared spectroscopy. The experimental results in Fig. 1e are almost in good agreement with the simulation, except for the slight blueshifts of the qBIC modes, which can be attributed to fabrication imperfections such as the shrinkage of the rod corners.

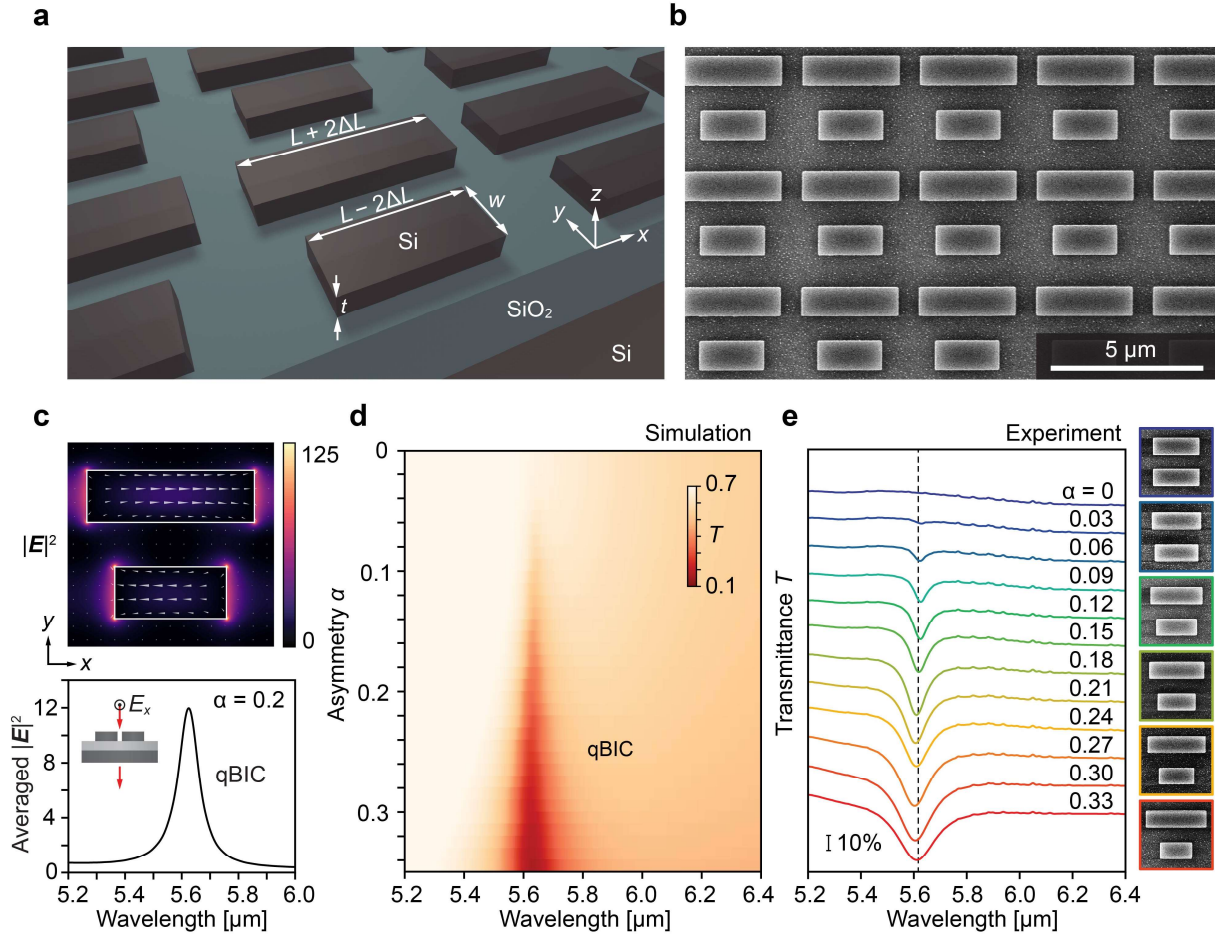

**Figure 1.** Silicon metasurfaces at qBICs. (a) Schematic of a silicon metasurface fabricated on an SOI wafer composed of two parallel asymmetric rods with an asymmetry parameter α = 2ΔL/L. (b) Representative SEM image of fabricated silicon metasurface (α = 0.2). (c) Simulated cross-sections of the $|E|^2$ profiles at the resonance peak wavelength (upper) and the spatially averaged $|E|^2$ spectrum (lower) in the xy-plane at half the height of the silicon metasurface (α = 0.2). The white cones indicate the displacement current distributions. (d) Simulated and (e) Experimental transmission spectra. The dashed lines indicate the approximate peak positions of the qBIC modes. The spectra are vertically shifted for clarity in (e). The qBIC modes were excited by vertically incident x-polarized light. The right panel in (e) shows the SEM images of the measured metasurface with α = 0, 0.06, 0.12, 0.18, 0.24, and 0.30 from top to bottom.

To analytically illustrate the experimental variations in the linewidths (= $\lambda_c/Q_{total}$, where $\lambda_c$ is the resonance mode of a qBIC mode) of metasurfaces with different α, the total Q factor $Q_{tot}$ is decomposed into the radiative ($Q_r$) and other nonradiative Q factors ($Q_{nr}$) as:

$$Q_{tot}^{-1} = Q_r^{-1} + Q_{nr}^{-1} = Q_r^{-1} + Q_{abs}^{-1} + Q_{scat}^{-1}. \tag{2}$$

Here, $Q_{abs}^{-1}$ and $Q_{scat}^{-1}$ are the material absorption and scattering losses due to surface roughness, respectively. To investigate the loss mechanisms produced by the asymmetric cladding layers of SOI wafers in the vertical direction, we calculate the Q factors and decompose the contributions from the three directions as follows [28, 29]:

$$Q^{-1} = \frac{P(t)}{\omega_0 U(t)} = \frac{P_{top}(t) + P_{bottom}(t) + P_{in}(t)}{\omega_0 U(t)} = Q_{top}^{-1} + Q_{bottom}^{-1} + Q_{in}^{-1}, \quad (3)$$

where $U(t)$ is the modal electromagnetic energy, and $P(t)$ is the radiation power absorbed in the calculation boundary. $P(t)$ is then separated into radiation in the in-plane direction ($P_{in}(t)$), upward direction ($P_{top}(t)$), and downward direction ($P_{bottom}(t)$), whose boundaries are positioned approximately at λ/4 above and below the silicon surface. In the calculation, Bloch boundary conditions are used in the $x$- and $y$- directions; therefore, the model is considered to have an infinite array size, and hence $Q_{in}^{-1} = 0$. In this case, $Q_r^{-1} = Q_{top}^{-1} + Q_{bottom}^{-1}$. The extinction coefficients $k$ of c-Si and SiO$_2$ are assumed to be zero for the calculations to separate $Q_r$ and $Q_{nr}$ (see Supporting Information S1 for a nonzero $k$). The calculated $Q$ factors superimposed on the experimental results are shown in Fig. 2. The $Q_{top}$ is well fitted by the well-known relationship that holds for the asymmetry parameter α added to the nanostructure [16]:

$$Q_r = Q_0 \alpha^{-2}, \quad (4)$$

The fitting line gives a constant value $Q_0$ of 22.7, signifying that the radiative component of the external medium, particularly in the upward direction of the metasurfaces, originates from the qBIC modes. Although SiO$_2$ has a small absorption around the resonance wavelength in our work [30], the relation $Q_{nr} > Q_{bottom}$ is fulfilled for all values of α. This implies that the limiting factor of the total $Q$ factors is neither the absorption loss nor the scattering loss, but the leakage loss into the bottom oxide layer of the SOI wafer [31]. This significantly limits the total $Q$ factors, especially for a small α. The existence of leakage losses is also evident from the field intensity for the cross-section in the $yz$-plane (Fig. 2b) at the center of the pair-rod. The total $Q$ factor, $Q_{tot}$, was calculated using Eq. (2), with $Q_{nr}$ as the fitting parameter, agreed well with the experimental $Q_{tot}$. This result suggests that our analytical model is sufficient for explaining the experimental $Q$ factors obtained for silicon metasurfaces using SOI wafers. It may be noted that the downward leaky losses can be reduced if the BOX layer is sufficiently thick to optically separate the confined qBIC mode from the bottom silicon substrate, as discussed later.

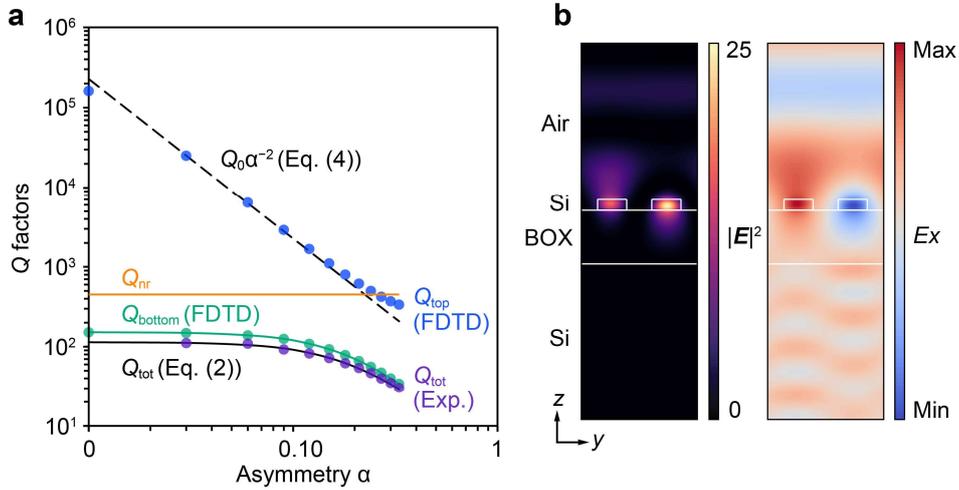

**Figure 2.** Effect of leakage losses into the high-index substrate of SOI wafer. (a) $Q$ factors of metasurfaces with different α. $Q_{top}$ (blue dots) and $Q_{bottom}$ (green line) represent $Q$ factors that contribute to upward and downward radiations, respectively, calculated by FDTD. $Q_{top}$ is fitted to Eq. (4) (dashed line). $Q_{nr}$ (orange line) is determined so that the experimental $Q_{tot}$ (purple dots) is fitted with the theoretical $Q_{tot}$ (black line), which is calculated by summing all the loss contributions above. (b) Simulated $|E|^2$ profile (left) and $E_x$ profile (right) in the $yz$-plane at the center of the unit cell for the metasurface with α = 0.2.

## 2. Coupling between qBIC modes and PMMA molecules

Next, a 111-nm PMMA layer was spin-coated onto a bare metasurface to characterize the coupling between the qBIC mode and the C=O stretching mode of PMMA. For the simulation, the dispersive dielectric constant of the PMMA molecules was modeled as a single-oscillator Lorentz permittivity whose parameters were experimentally determined using an infrared spectroscopic ellipsometer (see the Methods section). Figure 3a,b compares simulation and experimental spectra, respectively. Although there was a slight detuning between the qBIC modes and the C=O stretching modes of the PMMA absorption in the experiment, it was in good agreement with the simulation. Split peaks centered at the wavelengths of the C=O mode appeared when α was large. The amplitudes of the split modes increased with α; however, the split modes $\lambda_+$ and $\lambda_-$ were nearly constant (Fig. 3c). Note that the middle peak in the spectrum with a triple lineshape, especially when α is small, may be attributed to intrinsic uncoupled PMMA molecules and/or intermolecular interactions for high molecular density [32].

To explore the properties of the split modes in more detail, we considered the eigenvalues of the coupled qBIC-PMMA system, which gives two energy states as follows [3]:

$$\omega_\pm = \frac{\omega_c + \omega_m}{2} - \frac{i}{2}(\gamma_c + \gamma_m) \pm \frac{1}{2}\sqrt{4g^2 + \left[(\omega_c - \omega_m) - i(\gamma_c - \gamma_m)\right]^2}, \qquad (5)$$

where $\gamma_c$ and $\gamma_m$ are the loss rates of the qBIC and vibrational modes, respectively ($2\gamma_c$ and $2\gamma_m$ are the respective full width at half maximum). $g$ is the coupling rate between the qBIC and vibrational modes. The two energy states were evaluated by observing the anti-crossing behavior in structures with varied detuning (= $\omega_c - \omega_m$). Specifically, we fabricated PMMA-coated metasurfaces (α = 0.12) with scaling factors $S$ from 0.95 to 1.05, which allowed us to linearly tune the qBIC resonance wavelengths. Figure 3d-f compares the FDTD-simulated spectra, experimental spectra, and plots of the experimental peak positions, all of which showed good agreement. When there is no detuning ($\omega_m = \omega_c$), Rabi splitting Ω is expressed as follows:

$$\Omega = \omega_+ - \omega_- = \sqrt{4g^2 - (\gamma_c - \gamma_m)^2}. \qquad (6)$$

In the experiment, Ω = 53 meV was obtained, as shown in Fig. 3f, suggesting that the strong coupling condition $g > \gamma_m, \gamma_c$ was satisfied [33] when α = 0.12. Under this condition, the two newly formed polariton modes never cross each other, which is known as anti-crossing behavior. Comparing the almost constant $g$ and engineered $\gamma_c$ for different α, the rough estimate provided a strong coupling regime when α < 0.27 and weak coupling regime when α > 0.27. This result clearly shows that the coupling conditions can be tailored using the asymmetry parameter α of silicon metasurfaces. In an earlier study, changing the radiative $Q$ factors also affected the resonance wavelengths of metasurfaces [17]. In contrast, in our BIC metasurfaces, only the $Q$ factors were changed while maintaining the resonance wavelengths when the asymmetry parameters were changed. This provides an effective means of controlling the coupling conditions with vibrational transitions.

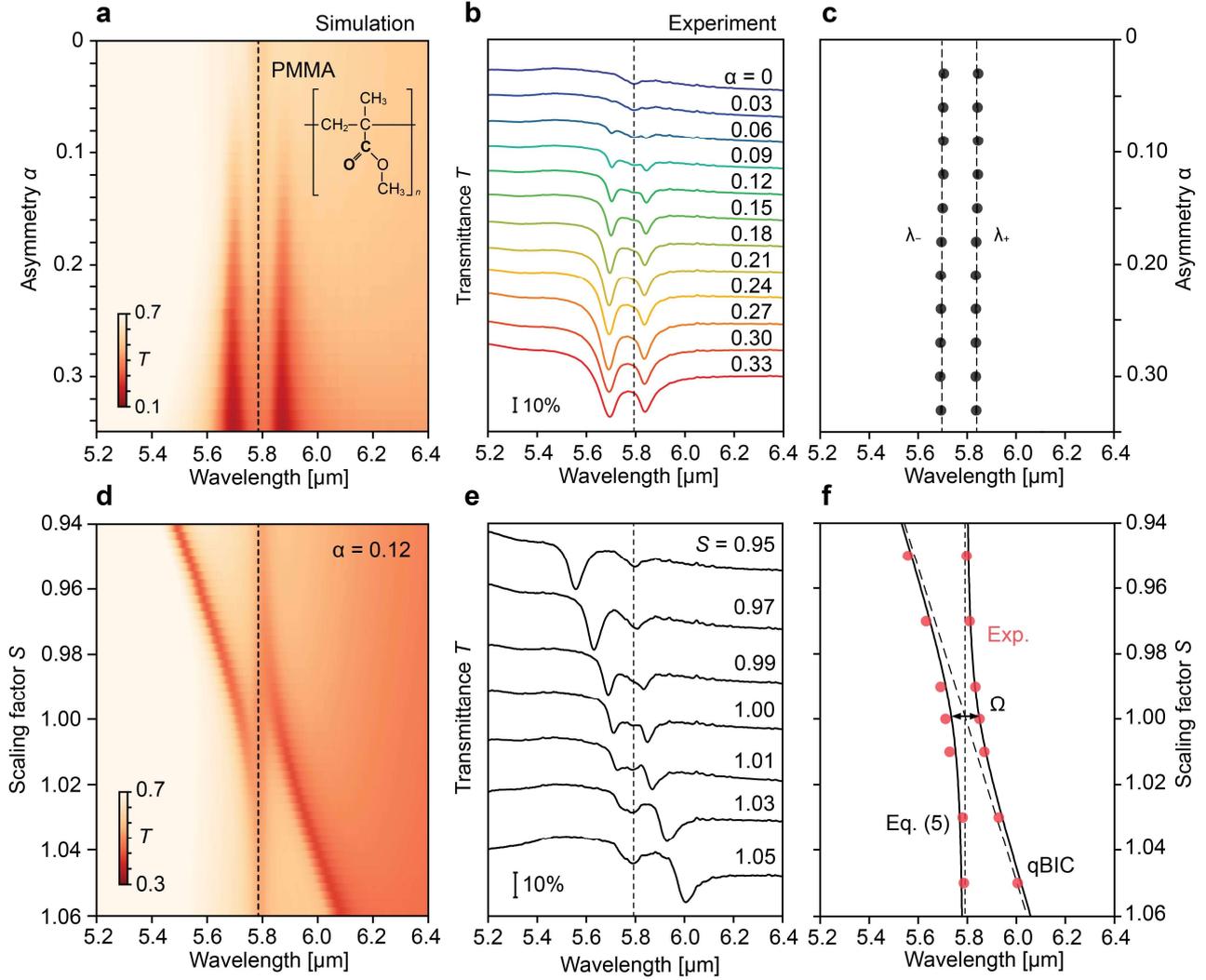

**Figure 3.** Coupling between qBIC modes in silicon metasurfaces and PMMA molecules. (a) Simulated transmission spectra of coupled qBIC-PMMA system with different α. Inset shows the chemical structure of PMMA. (b) Experimental transmission spectra, and (c) split modes $\lambda_+$ and $\lambda_-$ as functions of α. Dashed lines represent the respective average values. (d) Simulated anti-crossing behavior in the coupled system, obtained from the metasurfaces with different scaling factors $S$. (e) Experimental transmission spectra. (f) Experimental peak positions of the split modes (red dots) and the superimposed eigenvalues of Eq. (5). Ω represents Rabi splitting. The dashed line represents the peak wavelength of the qBIC modes, and the dotted line represents the absorption wavelength of the PMMA molecules. The spectra are vertically shifted for clarity in (b) and (e). The qBIC modes were excited by vertically incident $x$-polarized light.

## 3. Enhanced molecular signal

To predict the theoretical molecular signal strength and explore the dominant factor of surface enhancement effects in the coupled qBIC-PMMA system, we used the temporal coupled mode theory (TCMT) [34]. Figure 4a displays the TCMT model of a coupled cavity-molecular system with two input ports, namely, input traveling wave $s_+ = [s_{1+}\ s_{2+}]^T$ and output traveling wave $s_- = [s_{1-}\ s_{2-}]^T$. From the coupled mode equations, the transmission amplitude of the transverse electric (TE)-like mode can be derived (see Supporting Information S3) by considering the reflection coefficient $s_{1-}/s_{1+}$, as follows:

$$T = 1 - \left|\frac{s_{1-}}{s_{1+}}\right|^2 = 1 - \left|r - \frac{2e^{i2\theta_1}/\tau_1}{i(\omega-\omega_c) + \frac{1}{\tau_1} + \frac{1}{\tau_2} + \frac{1}{\tau_{nr}} + \frac{g^2}{i(\omega-\omega_m)+1/\tau_m}}\right|^2. \tag{7}$$

Here, we assume that the coupled system couples outside the medium with an upward radiative rate of $\tau_1^{-1} = \omega_c Q_{top}^{-1}/2$ and a downward radiative rate of $\tau_2^{-1} = \omega_c Q_{bottom}^{-1}/2$. The nonradiative rate $\tau_{nr}^{-1}$ accounts for material absorption and scattering losses as $\tau_{nr}^{-1} = \omega_c Q_{nr}^{-1}/2 = \omega_c(Q_{abs}^{-1} + Q_{scat}^{-1})/2$. $\theta_1$ is the phase related to the coupling coefficient between the two ports and resonance modes. In SOI wafers, the transmission spectrum has a broad Fabry–Perot (FP) background owing to FP resonances in the oxide layer. The interference between the broad background spectrum and the sharp qBIC mode explains the Fano lineshape in the qBIC modes, which is expressed by the reflection coefficient $r$ and transmission coefficient $t$ in Eq. (7). Next, we define the relative transmittance $T_r = T - T|_{\alpha=0}$ obtained by subtracting the transmittance for $\alpha = 0$, to exclude the influence of the middle peak of the triple peak appearing in the transmission of a coupled qBIC-PMMA mode, because Eq. (7) does not take this effect into consideration. We then defined the enhanced molecular signal [22, 23] as follows:

$$\Delta T = T_r - T_r|_{g=0}. \tag{8}$$

This expression represents the transmission difference with and without PMMA molecules at a frequency of $\omega_c$ as depicted in Fig. 4b. Here, the $T_r|_{g=0}$ spectra were redshifted by considering the increase in the refractive index of the cladding layer when the PMMA molecules were adsorbed on the metasurface (the extent of redshift was determined by FDTD simulation). Experimentally, $\Delta T$ was determined from the transmittance difference peak at a frequency $\omega_c$ (Fig. 4c). Figure 4d,e shows the experimental transmittance dip amplitude $T$ and the enhanced molecular signal $\Delta T$, superimposed on their theoretical models obtained by considering a fundamental bound on $\tau_1/\tau_2$ constrained by $(1 + r)/(1 − r)$ [35, 36]. Good agreement with the experimental results validates the theoretical models and shows that the dependencies of the radiative losses on α significantly change $T$ and $\Delta T$. Importantly, we can determine the existence of the optimum α, where the highest molecular vibrational signal is obtained. The optimum α was approximately 0.24 under the present conditions. It should be noted that we implicitly assumed here that the coupling strength $g$ does not depend on the asymmetry parameter α for the calculation of $\Delta T$ using TCMT formalism. This is reasonable considering that the energy splitting was almost independent of α, as can be seen in Fig. 3 b,c. This was also true for the FDTD simulations. The constant peak splitting indicates that the α-dependence of the enhanced molecular signal $\Delta T$ is not caused by changes in $g$ but by the radiative loss engineering of qBIC modes [32, 37].

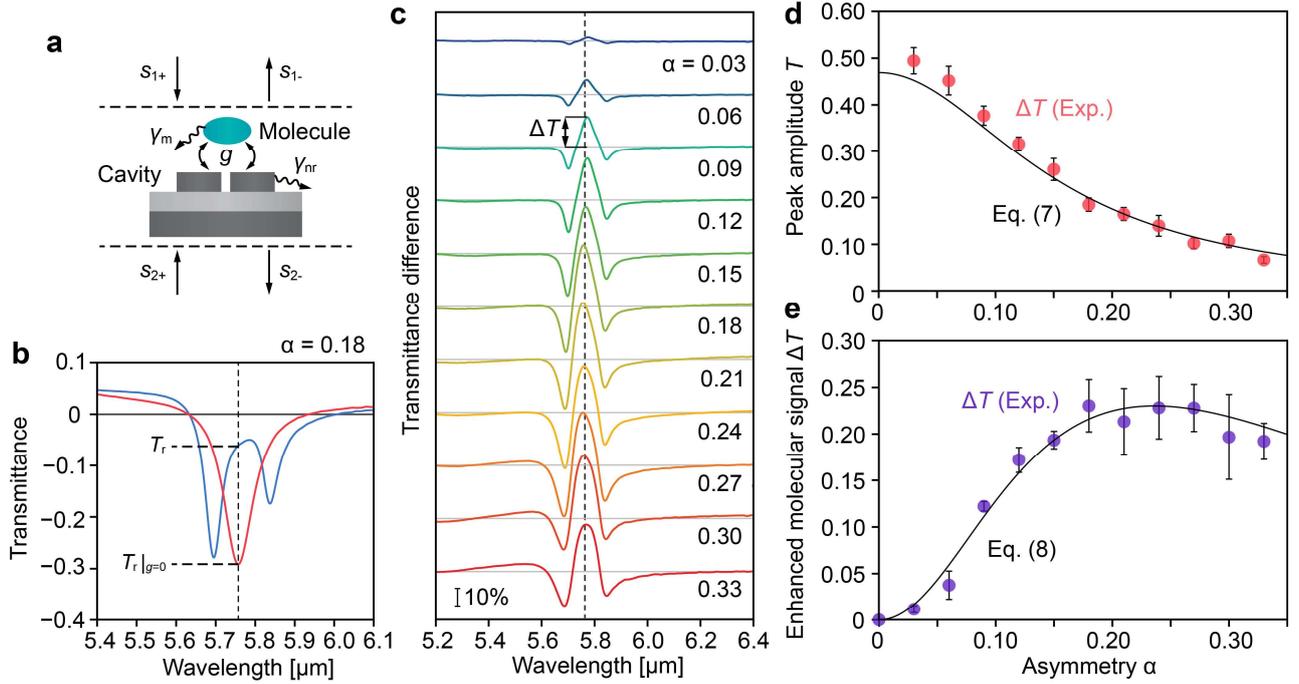

**Figure 4.** TCMT analysis and enhanced molecular signal of the coupled qBIC-PMMA system. (a) TCMT model of a two-port system interacting between a metasurface (cavity) and vibrating molecules. (b) Experimental relative transmittance spectra for α = 0.18. $T_r|_{g=0}$ and $T_r$ represent the transmittance amplitudes of the qBIC (red) and coupled qBIC-PMMA (blue) modes at the frequency $\omega_c$ (dashed line). (c) Transmittance difference spectra $T_r - T_r|_{g=0}$ for different α. For clarity, the spectra were shifted vertically. (d) Peak amplitudes $T$ of the qBIC modes for different α obtained experimentally (red dots) and the TCMT model (black line, Eq. (7)) (e) Enhanced molecular signal $\Delta T$ for different α obtained experimentally (purple dots) and the TCMT model (black line, Eq. (8)).

Finally, to explore the effect of the downward leaky losses, we calculate the enhanced molecular signal, $\Delta T$, for different BOX layer thicknesses. Figure. 5a shows the calculated $Q_{bottom}$, which is limited by the leaky losses toward the high-index substrate. Here, the scaling factors of the metasurfaces with different BOX thicknesses are slightly tuned such that the peak wavelengths of the qBIC modes and PMMA absorption match. As can be seen, the $Q_{bottom}$ becomes larger (= lower leaky losses) and follows the typical $\alpha^{-2}$ dependence for the qBIC modes when the BOX layer is thicker. Figure. 5b shows the calculated $\Delta T$. The optimum α at which the maximum $\Delta T$ is obtained is shifted to a smaller α, and the molecular signal strength increases when the BOX layer is thicker. These findings reveal the importance of controlling leaky losses and material choices. Specifically, a symmetric cladding with a small refractive index provides a higher molecular signal. However, a thick oxide layer produces unwanted FP resonances with a narrow peak wavelength spacing, which sometimes hinders weak qBIC and vibrational modes. Therefore, BOX layer thickness should be carefully designed depending on the resonance wavelength to be measured. Fortunately, the 2000-nm BOX layer in our work offered a moderate and sufficient background around the resonance wavelength of the qBIC mode (see Supporting Information S4 for more details), which provided a means for measuring the resonance dips and their enhanced molecular signals.

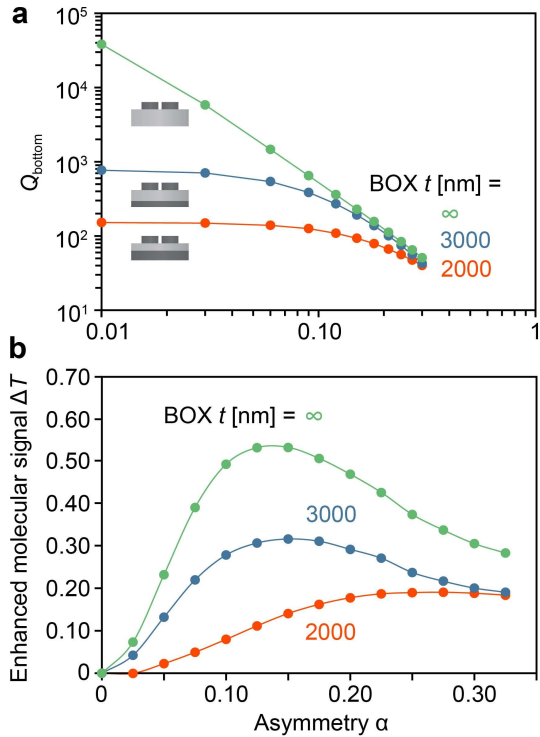

**Figure 5.** Effect of BOX layer thickness on (a) $Q_{bottom}$ originating from downward leaky losses and (b) enhanced molecular signal $\Delta T$ for different α. All the data points are calculated using the FDTD simulation, assuming nonzero material absorptions.

**Conclusions**

In this study, we have experimentally and theoretically performed vibrational coupling to the qBIC modes of silicon-based metasurfaces from weak to strong coupling regimes and demonstrated an enhanced molecular signal in infrared absorption. The coupling conditions were tuned by changing only the structural asymmetry parameters of the asymmetric pair-rod arrays, which were responsible for the changes in the radiative losses. We have shown the existence of an optimum asymmetry at which the highest molecular signal is obtained, indicating surface enhancement effects originating from the radiative loss engineering of silicon-based metasurfaces. The wide tunability of radiative losses in dielectric metasurfaces favors the optimization of the coupling conditions of the vibrational mode, opening up potential applications for highly sensitive surface-enhanced infrared spectroscopy without using metals. Although the total $Q$ factors of the coupled system are limited by leakage loss into the high-index bottom substrate, the use of SOI wafers allows the mass production of sensor chips, which is a significant advantage of using complementary metal-oxide semiconductor compatible dielectric materials. We believe that these findings can be applied to future research, including protein sensing [38], identification of small numbers of molecules [39], and modification of chemical reaction rates under strong coupling [40] using all-dielectric metasurfaces with optimized radiative losses.

**Materials and Methods**
**Fabrication**

Silicon metasurfaces were fabricated from an SOI wafer with 400-nm c-Si and 2000-nm $SiO_2$ layers. The SOI wafer was cleaned with acetone and isopropyl alcohol (IPA) for 10 min each in an ultrasonic bath. A positive resist (ZEP-520A, Zeon Chemicals) with a thickness of 100 nm was spin-coated at 6000 rpm for 60s onto the SOI wafer and prebaked on a hotplate at 180 °C for 3 min. Subsequently, a conductive layer (ESPACER 300Z) was spin coated at 2000 rpm for 60 s. A nanopattern (3 mm × 3 mm) was prepared using electron-beam lithography (ELS-BODEN, Elionix) at an acceleration voltage of 100 kV. The samples were then developed using xylene and IPA. Finally, the nanopatterns were transferred into the c-Si layer using silicon deep reactive ion etching equipment (MUC-21 ASE-SRE, Sumitomo Precision Products) with $SF_6$ and $C_4H_8$ gases (Bosch process). The remaining resist was removed via $O_2$ plasma ashing for 20 min.

**Characterization**

The transmission spectra of the 3 mm × 3 mm sample were acquired using a Fourier transform infrared spectrometer (Nicolet iS50R, Thermo Fisher Scientific) with a spectral resolution of 4 cm$^{-1}$ equipped with a liquid nitrogen-cooled mercury cadmium telluride detector. The polarization direction of the incident light was polarized in the *x*-direction using a wire-grid polarizer. To couple the PMMA molecules with the silicon metasurfaces, PMMA (950 A2) was spin-coated at 1500 rpm for 60 s onto the surface and baked on a hotplate at 180°C for 90 s. In determining the real and imaginary parts of the refractive index (*n*, *k*) of the PMMA molecules, an infrared spectroscopic ellipsometer (SENDIRA, SENTECH Instruments GmbH) was used. The absorption of the PMMA molecules was then modeled as a permittivity, represented by the Lorentz model [37]:

$$\varepsilon_{\text{PMMA}} = \varepsilon_\infty + \frac{f_0 \omega_0^2}{\omega_0^2 - \omega^2 - i\omega\Gamma} \tag{8}$$

where the background dielectric constant of PMMA molecules $\varepsilon_\infty$ = 2.20, Lorentz resonance frequency $\omega_0$ = 3.252 × 10$^{14}$ rad/s, strength coefficient $f_0$ = 0.016, and Lorentz damping rate $\Gamma$ = 3.41 × 10$^{12}$ rad/s.

**Simulation**

Transmission spectra, electric field distributions, and *Q* factors were acquired using a FDTD solver (ANSYS Lumerical). Bloch boundary conditions were used in the *x* and *y* directions, and perfectly matched layers were used in the ±*z* direction. The exact structural parameters were extracted from SEM images of the fabricated metasurface. The dielectric functions of Si and SiO$_2$ were taken from literature by Palik [41]. To simulate the coupled qBIC-PMMA system, the dielectric function of the C=O stretching mode of PMMA molecules was modeled using Eq. (8), which was experimentally determined using a spectroscopic ellipsometer, as described above.

**Acknowledgments:** We thank Kazuhiko Sakoda for technical support in the sample fabrication. The fabrication of the silicon metasurfaces was conducted at the Nanofabrication Platform and Namiki Foundry in NIMS.
**Author contributions:** All the authors have accepted responsibility for the entire content of this submitted manuscript and approved submission.
**Research funding:** This work was financially supported by JSPS KAKENHI Grant Number JP22K20496.
**Conflict of interest statement:** The authors declare no conflicts of interest regarding this article.

# Supporting Information for

**Vibrational coupling to quasi-bound states in the continuum under tailored coupling conditions**

Keisuke Watanabe[1]*, Hemam Rachna Devi[2], Masanobu Iwanaga[3], and Tadaaki Nagao[2,4]

1.* International Center for Young Scientists (ICYS), National Institute for Materials Science (NIMS), 1-1 Namiki, Tsukuba, Ibaraki 305-0044, Japan.
2. International Center for Materials Nanoarchitectonics (MANA), National Institute for Materials Science (NIMS), 1-1 Namiki, Tsukuba, Ibaraki 305-0044, Japan.
3. Research Center for Electronic and Optical Materials, National Institute for Materials Science (NIMS), 1-1 Namiki, Tsukuba, Ibaraki 305-0044, Japan.
4. Department of Condensed Matter Physics, Graduate School of Science, Hokkaido University, Kita 10, Nishi 8, Kita-ku, Sapporo 060-0810, Japan.

*E-mail: watanabe.keisuke@nims.go.jp

### S1. *Q* factors considering material absorptions

In the main text, the $Q$ factors were calculated with the extinction coefficient $k$ set to zero to separate $Q_r$ and $Q_{nr}$. Here, we calculate $Q$ factors considering the material absorptions of Si and SiO$_2$, whose dielectric functions were taken from Palik [41]. As shown in Fig. S1, the calculated $Q$ factors with and without material absorption exhibit similar trends with a slight difference. Because the total $Q$ factors with $k \neq 0$ are expressed as $Q_{tot}^{-1} = Q_R^{-1} + Q_{abs}^{-1}$, the $Q_{abs}$ is estimated to be approximately 900 in this wavelength range.

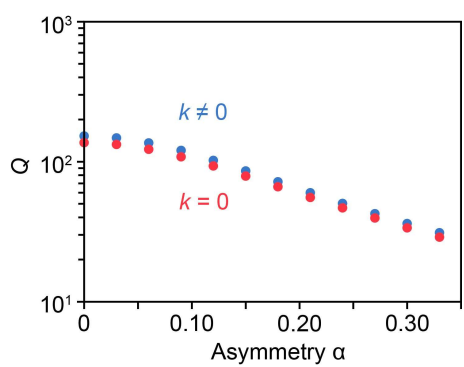

**Figure S1.** Comparison of the FDTD calculated $Q$ factors with and without consideration of material absorption.

### S2. Comparison of FDTD simulated and experimental spectra

As shown in Figure. S2, the simulation and experiment show good agreement for both the qBIC and coupled qBIC-PMMA resonances. The slight differences in both resonances indicate that the resonance peaks were slightly blueshifted in the experiment. This can be attributed to structural deformations (e.g., rounded corners) by fabrication errors. For coupled qBIC-PMMA resonances, the triple peaks are pronounced in the simulation, especially when α is small. An intuitive explanation for this is that the absorption of the intrinsic uncoupled PMMA molecules is unexpectedly large, presumably because of the less accurate dielectric constants of PMMA.

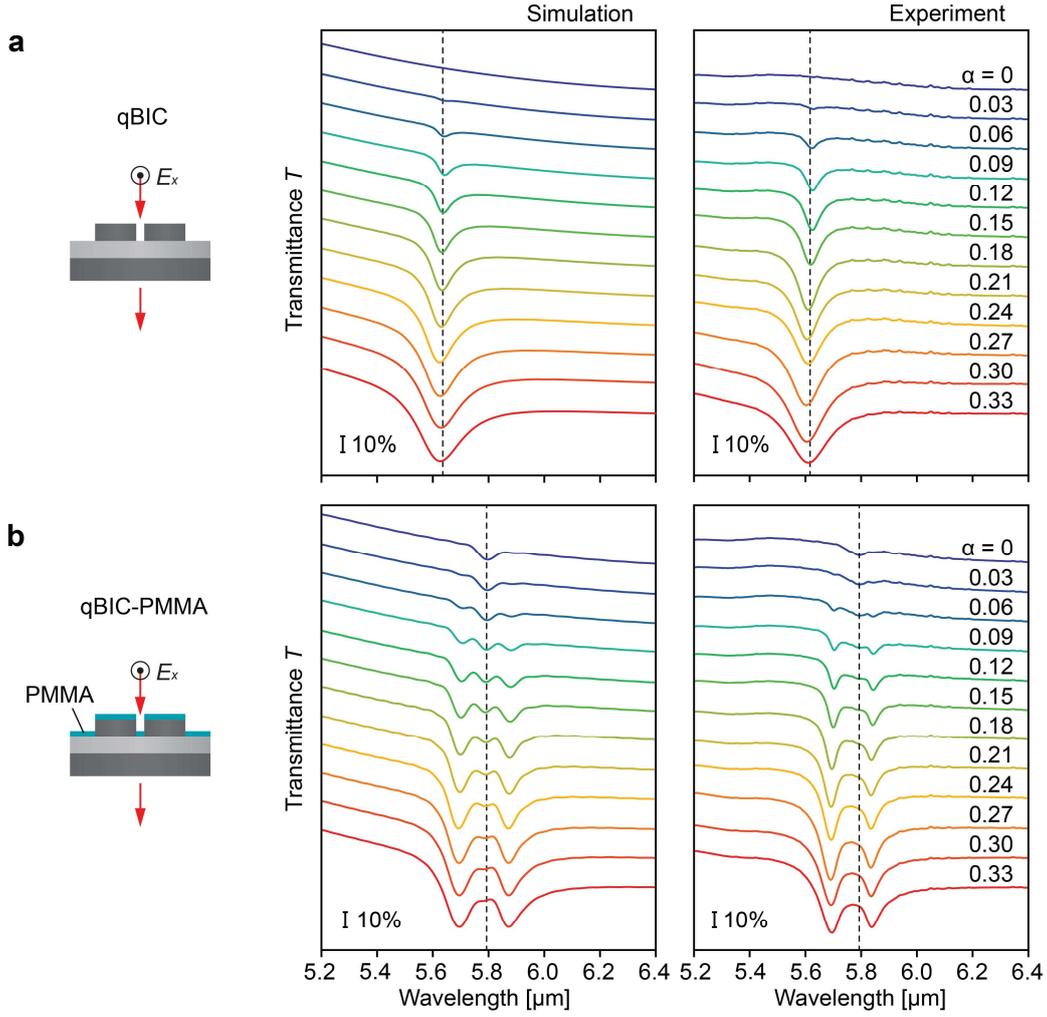

**Figure S2.** Transmission spectra of (a) bare and (b) PMMA-coated silicon metasurfaces with different α. The schematic shows a metasurface excited by vertically incident *x*-polarized light. The dashed lines indicate the approximate peak positions of the qBIC modes for (a) and the C=O modes for (b). For clarity, the spectra are shifted vertically.

### S3. Temporal coupled mode theory for asymmetric metasurfaces coupled with molecules

Temporal coupled mode theory (TCMT) offers a means of describing the resonance behaviors of a coupled cavity-molecular system as well as its enhanced molecular signal. In the following section, we derive a theoretical model for metasurfaces with asymmetric cladding layers coupled to molecules on the top surface. As illustrated in Fig. 4a in the main text, we consider a coupled cavity-molecular system with two input ports: the input traveling wave $s_+ = [s_{1+}\ s_{2+}]^T$ and the output traveling wave $s_- = [s_{1-}\ s_{2-}]^T$. The coupled-mode equations are as follows:

$$\frac{d}{dt}A = i\omega_c A - \left(\frac{1}{\tau_1} + \frac{1}{\tau_2} + \frac{1}{\tau_{nr}}\right)A + Ds_+ + igM, \tag{S1}$$

$$\frac{d}{dt}M = i\omega_m M - \frac{1}{\tau_m}M + igA, \tag{S2}$$

$$s_- = Cs_+ + D^T A, \tag{S3}$$

where $A$ and $M$ are the field amplitudes of the qBIC and vibrational modes, whose resonance frequencies are $\omega_c$ and $\omega_m$, respectively. The coupled cavity-molecular system has a coupling strength $g$, and the coupled system couples outside the medium with an upward radiative rate $\tau_1^{-1}$ and a downward radiative rate $\tau_2^{-1}$. $\tau_{nr}^{-1}$

is the nonradiave decay rate caused by the absorption of the materials and scattering losses due to fabrication errors. *C* is the scattering matrix for transmission and reflection without the resonances given by $C = e^{i\phi}\begin{pmatrix} r & it \\ it & r \end{pmatrix}$, where *r* and *t* are the reflection and transmission coefficients, respectively, that account for the Fano lineshape of the resonance caused by the interference between the broad background spectrum and the sharp cavity mode. ϕ is the phase factor, which depends on the reference plane position and is set to zero. *D* is the coupling coefficient between the two ports and is given by $D = (d_1\ d_2)$. Assuming that the resonance amplitudes *A* and *M* have time dependence $e^{-i\omega t}$ and $s_{2+} = 0$ (only one input port from the top), solving Eq. (S1)–(S3) gives an expression for the reflection spectrum of the transverse electric (TE)-like mode as follows:

$$R = \left|\frac{s_{1-}}{s_{1+}}\right|^2 = \left| r - \frac{2e^{i2\theta_1}/\tau_1}{i(\omega-\omega_c) + \frac{1}{\tau_1} + \frac{1}{\tau_2} + \frac{1}{\tau_{nr}} + \frac{g^2}{i(\omega-\omega_m) + 1/\tau_m}} \right|^2$$

$$= \left| r - \frac{2(\cos 2\theta + i \sin 2\theta)/\tau_1}{i(\omega-\omega_c) + \frac{1}{\tau_1} + \frac{1}{\tau_2} + \frac{1}{\tau_{nr}} + \frac{g^2}{i(\omega-\omega_m) + 1/\tau_m}} \right|^2.$$

(S4)

As mentioned in the main text, this equation does not consider the interactions between the molecules and the input port for simplicity (there is an interaction only between the cavity and molecules). Here, we assume $d_1 = \sqrt{\frac{2}{\tau_1}}e^{i\theta_1}$ and $d_2 = \sqrt{\frac{2}{\tau_2}}e^{i\theta_2}$ by defining the respective phases θ₁ and θ₂. From the energy conservation and time-reversal symmetry, *C* and *D* satisfy $d_1^*d_1 = 2/\tau_1$, $d_2^*d_2 = 2/\tau_2$, and $C\begin{pmatrix} d_1 \\ d_2 \end{pmatrix}^* = -\begin{pmatrix} d_1 \\ d_2 \end{pmatrix}$. By solving these equations, we obtain:

$$\cos 2\theta = \frac{\tau_1}{2r}\left(-\frac{r^2}{\tau} - \frac{1}{\sigma}\right),$$

$$\sin 2\theta = \pm \frac{\tau_1}{2r}\sqrt{\frac{4r^2}{\tau_1^2} - \frac{r^4}{\tau^2} - \frac{1}{\sigma^2} - \frac{2r^2}{\tau\sigma}},$$

(S5)

where

$$\frac{1}{\tau} = \frac{1}{\tau_1} + \frac{1}{\tau_2},$$

$$\frac{1}{\sigma} = \frac{1}{\tau_1} - \frac{1}{\tau_2}.$$

(S6)

The transmission spectra $T = 1 - R$ can be obtained by substituting Eq. (S5) and (S6) into (S4). As previously pointed out [35, 36], there is a fundamental bound on $\tau_1/\tau_2$ constrained by $(1 + r)/(1 - r)$. For the model fitting of the experimental data and calculation of the enhanced molecular signal, we also considered this constraint and used the FDTD-calculated $Q_{top}$ and $Q_{bottom}$.

**S4. Effect of BOX layer thickness on resonance spectra**

The presence of the BOX layer in an SOI wafer causes interference between the top silicon metasurface and bottom silicon substrate. The resulting Fabry–Pérot background sometimes hinders the weak qBIC and vibrational modes. Therefore, the BOX layer thickness should be carefully designed depending on the resonance wavelength to be measured. Figure S3 shows the transmission spectra of metasurfaces with different BOX layer thicknesses. When α = 0.1, the qBIC mode of interest appears, and the resonance dip becomes apparent when the background spectrum is close to 1, that is, when the BOX layer thickness is approximately 2500 nm. Nevertheless, the 2000-nm BOX layer in our work still provides a sharp resonance dip, providing a means for measuring the resonance modes.

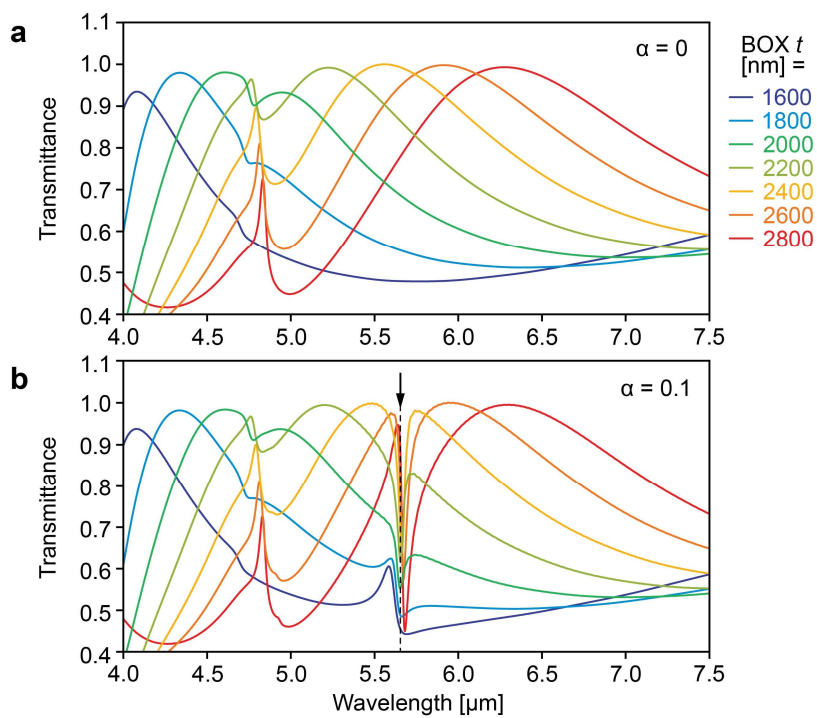

**Figure S3.** Transmission spectra of the metasurfaces with different BOX layer thicknesses when (a) α = 0 and (b) α = 0.1. The arrow indicates the wavelength at which the qBIC mode appears.